# The Evolution of Multicomponent Systems at High Pressures: II. The Alder-Wainwright, High-Density, Gas-Solid Phase Transition of the Hard-Sphere Fluid.

**J. F. Kenney**
*Joint Institute of the Physics of the Earth, Russian Academy of Sciences*;
*Gas Resources Corporation,*
11811 North Freeway, fl. 5, Houston, TX 77060, U.S.A.; JFK@alum.MIT.edu.?

Abstract:

The thermodynamic stability of the hard-sphere gas has been examined, using the formalism of scaled particle theory [SPT], and by applying explicitly the conditions of stability required by both the second and third laws of thermodynamics. The temperature and volume limits to the validity of SPT have also been examined. It is demonstrated that scaled particle theory predicts absolute limits to the stability of the fluid phase of the hard-sphere system, at all temperatures within its range of validity. Because scaled particle theory describes fluids equally well as dilute gases or dense liquids, the limits set upon the system stability by SPT must represent limits for the existence of the fluid phase and transition to the solid. The reduced density at the stability limits determined by SPT is shown to agree exactly with those of that estimated for the Alder-Wainwright, supercritical, high-density gas-solid phase transition in a hard-sphere system, at a specific temperature, and closely over a range of more than 1,000K. The temperature dependence of the gas-solid phase stability limits has been examined over the range 0.01K-10,000K. It is further shown that SPT describes correctly the variation of the entropy of a hard-core fluid at low temperatures, requiring its entropy to vanish as $T \to 0$ by undergoing a gas-solid phase transition at finite temperature and all pressures.[†]

## 1. Introduction.

The high-pressure, supercritical, gas-liquid phase transition was first reported specifically for the real fluid normal octane, $n\text{-}C_8H_{18}$, in a previous article which will be referred to henceforth as I.[1] That particular molecular specie had been chosen for analysis, perhaps fortuitously, for reasons of an availability of data. In I, it was asserted that the high-pressure, supercritical, first-order, gas-liquid phase transition is a geometric property of fluids which must exist independently of the details of any attractive component of the intermolecular potential. This argument was supported in I by showing that the gas-liquid phase transition would occur in a system composed of imaginary "isomers" of n-octane which, while maintaining all other properties of that fluid, might possess an SPHCT characteristic temperature of one millikelvin, $T^* = 0.001K$. Such a characteristic temperature would describe a fluid effectively without any long-range, attractive van der Waals component to its intermolecular potential, and which would be a gas at all temperatures and low pressures. That "isomer" of n-octane was shown to undergo the gas-liquid phase transition at almost the same pressure as the real fluid.

---

[†] [Keywords: theory, phase transitions, scaled particle theory, gas-solid, thermodynamics.]

In order to demonstrate rigorously that the high-pressure, supercritical, first-order, gas-liquid phase transition is a geometric property of fluids, in this article and a following one, the stability of hard-core fluids is explicitly investigated. Because the formalism of scaled particle theory [SPT] was applied extensively in I as the reference system for the joint partition function, the competence of that formalism to describe the stability of a hard-core system has been investigated thoroughly. In both this and the article to follow, the ability of SPT to predict phase transitions in a hard-core system is examined by applying all the conditions of stability required by the three laws of thermodynamics.

The purpose of this article is to demonstrate the ability of scaled particle theory to predict the Alder-Wainwright, high-density, gas-solid phase transition in a hard-sphere gas. There has been common misunderstanding that SPT only reproduces the pressure-density relationship of a hard-sphere system from low densities until, and through, the Alder-Wainwright phase transition, but fails to predict or reproduce that transition. Such misunderstanding is corrected in the following sections where is explicitly described the prediction of the Alder-Wainwright transition by SPT when constrained by general stability conditions. The density at which the fluid-solid phase transition occurs in a hard-sphere system is shown to depend upon its temperature. Furthermore, the temperature-dependence of the fluid-solid phase transition predicted by SPT requires that the fluid undergo such transition *at all pressures* as the temperature decreases to very low temperatures, precisely in keeping with general thermodynamic considerations. Additionally, the latent change of entropy of the fluid-solid transition predicted by SPT for hard sphere systems at low temperatures is shown to be consistent with experimental values of the melting entropy for volatile spherical molecules. In the following article are described the geometric effects of shape, degree of clustering, and asphericity upon the gas-liquid hard-core phase transition.

Because scaled particle theory uses a factored partition function which obtains from application of the Baker-Campbell-Hausdorff expansion, it is a high-temperature formalism. Therefore, the low-temperature limits of the validity of SPT have been examined explicitly and are described below.

The stability of the one-dimensional, "hard-rod," and two-dimensional, "hard-disk" systems have also been investigated and are described in two Appendices. When the general stability conditions, as used with the three-dimensional, hard-sphere fluid, are applied to the one-dimensional, "hard-rod" system, scaled particle theory predicts no instability against a fluid-solid phase transition of the Alder-Wainwright type. When those same stability conditions are applied to the two-dimensional, "hard-disk" system, it is shown to manifest a fluid-solid, "freezing" transition of the Alder-Wainwright type at reduced densities somewhat greater than 0.8.

## 2. Background.

The hard-core fluid has been extensively studied, especially since the discovery of the supercritical, high-density, gas-*solid* phase transition by Alder and Wainwright in the hard-sphere system.[2] A phase transition in a hard-sphere system had been predicted earlier in the theoretical work by Kirkwood, who argued that an ordered configuration should appear in the system at a reduced density of approximately 0.67.[3] Later, Born and Green developed the same conclusions.[4] However, the investigations by Kirkwood (and also those by Born and Green) involved extrapolations of an assumed two-particle correlation function for dense fluids from its form in the thermodynamic regime of dilute gases. The calculations of Kirkwood (and Born) could not predict accurately even the fourth virial coefficient for simple fluids; and such failure cast severe doubts about the validity of their predictions of a phase transition. The explicit solution by Alder and Wainwright of the equations of motion of assemblages of hard spheres in 1957, and their demonstration of the phase transition, removed all uncertainties about its existence.

During the same period while Alder and Wainwright were conducting their computer experiments, Reiss, Frisch, and Lebowitz were developing a formalism for the analysis of hard-body systems which applied a radically different perspective to statistical mechanics and employed a mathematical expression of the arcane branch of mathematics designated *statistical geometry*.[5] Their formalism would be named scaled particle theory [SPT]. The description of the thermodynamic properties of the hard-sphere gas by scaled particle theory represents one of the few exactly-solvable problems in quantum statistical mechanics. The resolution of the hard-sphere gas problem was later obtained, independently, by also Wertheim[6] and Thiele[7] who used different methods and whose results confirmed those developed first by Reiss, Frisch, and Lebowitz using SPT. (The chronology of the solution of the hard-sphere problem was stated incorrectly in I and is herewith corrected.)

Contemporaneously with the development of scaled particle theory by Reiss, Frisch, and Lebowitz, statistical geometry was being applied also by Bernal to experimental investigations of simple fluids.[8-11] Bernal first introduced explicitly the term statistical geometry and referred in his own work to the (relatively scarce) investigations in pure mathematics of that subject published by Kirkman[12,13], Cayley[14], Cundy[15], Toth[16], and Coxeter[17].

The connection between statistical mechanics and statistical geometry had been recognized by Boltzmann[18]. The reduction of statistical thermodynamic problems of dense fluids to ones of statistical geometry was taken up later by Reiss[19], and by Speedy[20,21], and still later by Reiss and Speedy together[22]. However, in the papers by Reiss, there is no reference to the work involving statistical geometry by Bernal, nor vice versa; neither seems to have been aware of the work of the other.

A review of many investigations of the hard-sphere system undertaken during the decade following the enunciation of its high-density, gas-solid phase transition by Alder and Wainwright is given in the book edited by Temperley *et al.* in 1968.[23] Notwithstanding the intended thoroughness of that review, there is nowhere in any of the many articles and hundreds of references contained therein even a *single mention* of scaled particle theory, nine years after its publication. Such was a staggering oversight.

The plethora of publications dealing with the hard-sphere gas has almost certainly been responsible for the series of coincidences which lay behind the oversight of direct application of scaled particle theory to the problems of the gas-solid and gas-liquid phase transitions in hard-core systems. That oversight has occasioned this article.

## 3. The thermodynamic conditions for stability.

Each of the three laws of thermodynamics states different conditions for the stability of a system. All must be satisfied for the stability of a system. The first law, being a statement of balance, imposes the obvious condition of stability which requires that, with no input of heat or mechanical work from the rest of the universe, an otherwise stable system remains unchanged. The second law, being a statement of irreversibility, imposes the more stringent condition of stability that, for any spontaneous transition, the *internal* component of the entropy of the system must *increase*. The third law, being a statement of the absolute property of the entropy, imposes the condition, upon not only the stability but also the existence of the system, that in any state its total entropy must be positive.[24-27]

The first and second laws together develop the conditions for stability of a single state of a single-component, single-phase system which require that its isothermal compressibility be greater than zero, that its isometric specific heat be greater than zero, that its isobaric specific heat be greater than its isometric specific heat, that the first derivative of its chemical potential with respect to pressure be positive and the second derivative be negative:

$$\left. \begin{array}{ll} -\dfrac{1}{V}\left(\dfrac{\partial V}{\partial p}\right)_T > 0 & \left(\dfrac{\partial p}{\partial T}\right)_V \left(\dfrac{\partial T}{\partial V}\right)_p > 0 \\[6pt] T\left(\dfrac{\partial S}{\partial T}\right)_V = C_V > 0 & \left(\dfrac{\partial \boldsymbol{m}}{\partial p}\right)_{T,n} > 0 \\[6pt] T\left(\dfrac{\partial S}{\partial T}\right)_p = C_p > C_V & \left(\dfrac{\partial^2 \boldsymbol{m}}{\partial p^2}\right)_{T,n} < 0 \end{array} \right\} \qquad (1)$$

The conditions for stability set out in equations (1) are concerned with any single state of a given system. As such, these conditions are *local*, a property evidenced by their expression as differential inequalities.

The second law constitutes a *global* constraint upon the stability of a system, as evidenced in its statement by De Donder's inequality:

$$A(S,T,n) = -\sum_{i,\boldsymbol{f}} \boldsymbol{n}_i^{\boldsymbol{f}} \boldsymbol{m}_i^{\boldsymbol{f}} > 0 \qquad \forall \{i, \boldsymbol{f} \ni B\} \qquad (2)$$

In equation (2), the summations are taken over all phases, $\boldsymbol{f}$, and all other possible final states, $\{``B"\}$, of the system; the mathematical entity, $A(S,T,n)$, is the thermodynamic Affinity, so defined by De Donder.[28] The non-local condition set forth by (2) places constraints upon the possibility that a system in one state might evolve spontaneously into another. Let the state considered stable be designated "*A*," and that of any other state "*B*." The second law, De Donder's inequality, requires explicitly that, for the stability of the system in state "*A*" against spontaneous transformation to any other state "*B*," i.e., "*A*" → "*B*," such hypothetical transformation must involve an *increase* in the internal entropy of the system. It deserves to be noted that, while De Donder's inequality sets a necessary condition for stability, it sets the sufficient condition for phase separation only when the states involved refer unequivocally to separated phases. Pertinent to the investigation of the ability of scaled particle theory to predict the Alder-Wainwright, high-density, gas-solid phase transition, it deserves note that, as was pointed out in I, previous analyses of high-density systems had failed to include an adequate set of trial states in their partition functions, and to apply the compulsions of De Donder's inequality.

The third law also sets additionally a stringent condition for the physical existence of any state: The total entropy of the system must be positive:

$$S \geq 0. \qquad (3)$$

The inequality (3) is, in fact, only a weaker corollary of the third law, and does not address the behavior of the entropy as temperature approaches absolute zero. Strictly, inequality (3) may be considered a consequence of Boltzmann's classical definition:
$$S = k \ln W, \qquad (4)$$
in which $W$ represents the number of configurations which the system can assume, consistent with its thermodynamic constraints. Because $W$ is a natural number, $\geq 1$, the inequality (3) results. The enunciation of the third law by Nernst in 1906[29] demonstrated extraordinary prescience. That the entropy of a system goes to zero with temperature is now attributed to the limitations of the degeneracy of the ground states of systems at the quantal level.[30] The third law states that entropy is an absolute entity, *with no additive constants*. The absolute property of entropy is now recognized as a consequence of the formalism of general relativity. When cast into fully covariant form as required by general relativity, the formalism of thermodynamics admits no affine extensive functions.[31] Unlike De Donder's inequality, a failure of Nernst's theorem,[29,32], for any phase of a system constitutes both a necessary and sufficient condition for a phase transition.

## 4. Scaled particle theory [SPT].

The scaled particle theory developed by Reiss, Lebowitz and Frisch[5,33-40], as extended by Carnahan and Starling[41], generates the explicit canonical partition function:

$$\Omega(V,T;N)^{SPT} = \Omega^{IG} \cdot \Omega^{h-c} = \frac{1}{N!}\left(\frac{V}{\boldsymbol{l}^3}\right)^N \cdot \left[\exp\left(-c\frac{\boldsymbol{h}(4-3\boldsymbol{h})}{(1-\boldsymbol{h})^2}\right)\right]^N. \qquad (5)$$

In equation (5), the "thermal de Broglie wavelength," $\boldsymbol{l} = h/\sqrt{(2\pi m k_B T)}$, where $h$ is Planck's constant and $k_B$ Boltzmann's constant; $T$ is the absolute temperature and $m$ the molecular mass; $\boldsymbol{h}$ is the "free-volume" parameter, $\boldsymbol{h} = \boldsymbol{t}V^*/V = \boldsymbol{t}n\mathrm{v}^*/V$, in which $\boldsymbol{t}$ is the close-packing parameter $\sqrt{2}\pi/6 \approx 0.7405$, $n$ the molar abundance, and $\mathrm{v}^*$ the molar covolume, or excluded volume. In equation (5), the exponent in the factor for hard-core contribution includes the Prigogine $c$-factor to account for the asphericity and possible flexibility of the molecule; in a hard-sphere gas, $c = 1$.[42] It is emphasized that the two parameters in the SPT canonical partition function, $V^*$, and $c$, are *not* adjustable parameters; the values of $V^*$ and $c$ are determined independently.

The factored partition function, (5), constitutes implicit application of the Baker-Campbell-Hausdorff lemma,

$$Q = tr\langle\exp(-\boldsymbol{b}H(p,q))\rangle = tr\langle\exp(-\boldsymbol{b}(T(p)+\Omega(q)))\rangle$$
$$\neq tr\langle\exp(-\boldsymbol{b}T(p))\rangle tr\langle\exp(-\boldsymbol{b}\Omega(q))\rangle \qquad (6)$$
$$= tr\langle\exp(-\boldsymbol{b}T(p))\rangle tr\langle\exp(-\boldsymbol{b}\Omega(q))\rangle tr\left\langle\left(1-\frac{\boldsymbol{b}^2}{2}[T,\Omega]+\cdots\right)\right\rangle$$

Neglect of the third factor on the right side of (6) constitutes a high-temperature approximation. Thus, from the form of its partition function, SPT is a formalism valid only at high temperatures. The low temperature limits for the validity of the SPT formalism are investigated explicitly in Section 6.

The SPT canonical partition function generates the Helmholtz free energy:

$$F^{\text{SPT}} = N\text{k}_\text{B}T\left[-\ln\left(\frac{V}{N\boldsymbol{l}^3}\right)+1+c\frac{\boldsymbol{h}(4-3\boldsymbol{h})}{(1-\boldsymbol{h})^2}\right]. \qquad (7)$$

The volume derivative of the logarithm of the SPT canonical partition function gives the system pressure:

$$p = N\text{k}_\text{B}T\left[\frac{1}{V}+\frac{c}{V}\cdot\frac{2\boldsymbol{h}(2-\boldsymbol{h})}{(1-\boldsymbol{h})^3}\right] = p^{\text{IG}}+p^{\text{hc}}. \qquad (8)$$

In equation (8), the two terms may be considered as, respectively, the "ideal gas" term and the hard-core term:

$$\left.\begin{array}{l}p^{\text{IG}} = N\text{k}_\text{B}T\dfrac{1}{V} \\[1em] p^{\text{hc}} = N\text{k}_\text{B}T\dfrac{c}{V}\dfrac{2\boldsymbol{h}(2-\boldsymbol{h})}{(1-\boldsymbol{h})^3}\end{array}\right\}. \qquad (9)$$

An important aspect of scaled particle theory, which obtains from its property as an exact, geometric description, is that it generates an absolute solution. There are no adjustable parameters or additive constants to the solution to the problem of the hard-core gas given by scaled particle theory. This property of scaled particle theory was stated succinctly by one of its principal authors more than two decades ago: "*A particularly dramatic feature of the theory* [SPT] *is the fact that it contains no adjustable parameters, and that the numbers which agree with experiment are obtained from formulae containing quantities which are derived from independent measurements.*" [Emphasis set forth in the original.][35] This property of scaled particle theory is used directly below to predict, unequivocally, the [Alder-Wainwright] high-density, gas-*solid* phase transition, both at high temperatures and as the temperature of the system approaches absolute zero. The same property is used also in the following article to predict the [Kenney] high-density, gas-*liquid* phase transition and to resolve the determining factors involving asphericity and the degree of clustering.

## 5. The absolute entropy of a hard-sphere system, and the Alder-Wainwright, high-density, gas-solid phase transition.

The hard-core contribution to the partition function as described by scaled particle theory is purely entropic. This property results intrinsically from the arguments from geometric statistics which underlie scaled particle theory and are indifferent to temperature. Therefore, the only contribution to the internal energy of a hard-sphere fluid obtains from the ideal gas factor in its partition function, such that,

$$U^{SPT} = \frac{3}{2} N k_B T . \tag{10}$$

When the general thermodynamic equation for the Helmholtz free energy,
$$F = U - TS , \tag{11}$$
is applied, the entropy for a hard-core gas is given explicitly by the Carnahan-Starling extension of scaled particle theory as:

$$S^{SPT} = N k_B \left( \frac{5}{2} + \ln\left( \frac{V}{n N_A \mathbf{l}^3} \right) - c \cdot \frac{\mathbf{h}(4 - 3\mathbf{h})}{(1 - \mathbf{h})^2} \right). \tag{12}$$

In equation (12), the first two terms may be recognized as, effectively, the "ideal gas" terms and the last the hard-core term.

$$\left. \begin{array}{l} S^{IG} = N k_B \left( \dfrac{5}{2} + \ln\left( \dfrac{V}{n N_A \mathbf{l}^3} \right) \right) \\[1em] S^{hc} = -N k_B c \cdot \dfrac{\mathbf{h}(4 - 3\mathbf{h})}{(1 - \mathbf{h})^2} \end{array} \right\}. \tag{13}$$

The first of equations (13) is recognized as the Sakur-Tetrode equation, to which the SPT equation for the entropy reduces in circumstances of large volumes and dilute densities. Although the so-called ideal gas contribution to the entropy, $S^{IG}$ the first of equations (13), has the identical functional form as the Sakur-Tetrode equation for an ideal gas, the volume which appears in that function is related to the temperature and pressure by the SPT equation of state, (8), not by the ideal gas law, the first of equations (9).

The two leading terms in the entropy, the ideal gas contribution, $S^{IG}$, are always positive, and dominate that function in the region of low and moderate densities. The third term in the entropy, the hard-core contribution, $S^{hc}$, is always negative and, most importantly, possesses a second-order singularity in the density. Therefore, at sufficiently high densities, the entropy of the hard-core fluid must change sign and become negative.

Such behavior by the system would constitute a glaring violation of the third law of thermodynamics. Therefore, the density at which the entropy of the hard-core fluid changes sign represents an absolute limit *for the fluid*,
which cannot exist at that, or higher, densities. This limit determines the Alder-Wainwright, high-density, gas-solid phase transition. Scaled particle theory describes equally the dilute gas and the dense liquid states. Therefore, the limit determined by scaled particle theory for the existence of the system so described must set the boundary for a solid phase.

This property of the entropy of the hard-core fluid is shown graphically in Fig. 1 for a hard-sphere gas, for which $c = 1$, at the temperature 300 K, as a function of its reduced density, ***h***. In Fig. 1, are shown graphically also the respective contributions of the ideal gas and hard-core entropies. Also in Fig. 1 is drawn the vertical line which represents the reduced density for the gas-solid phase transition as estimated by Alder and Wainwright using the fixed value for the free-volume parameter, ***h***$^{AW}$ = 0.67114.[43] The position of the gas-solid phase transition estimated by the fixed value of the free-volume parameter, ***h***$^{AW}$, is seen to lie very closely, within a few percent, to that determined by the vanishing of the SPT entropy. In fact and as may be observed in Fig. 1, the estimate of the Alder-Wainwright transition using the fixed value, ***h***$^{AW}$ would allow the fluid, at least at the temperature 300K, to possess a negative entropy within the range of pressures between that determined by the sign change of the entropy and ***h***$^{AW}$.

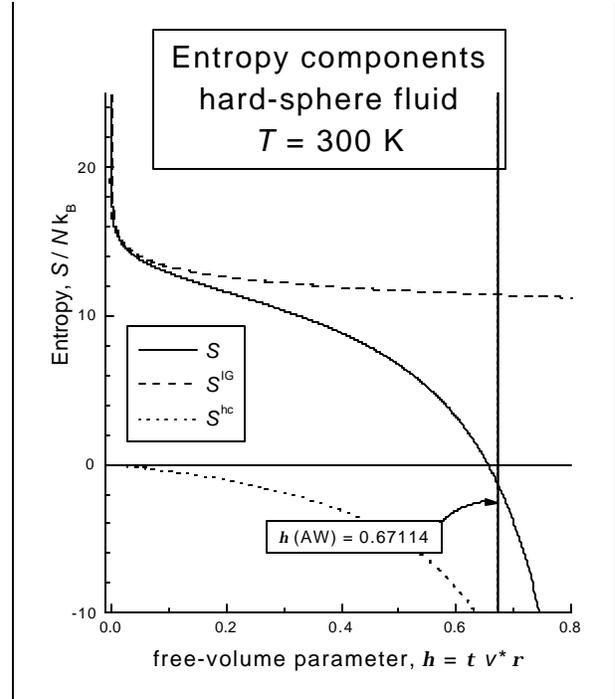

**Fig. 1 Entropy of a hard-sphere gas at 300 K.**

The inadequacies of the local stability conditions set out in equation (1) to predict the Alder-Wainwright transition may be noted by calculation of the isothermal compressibility, $-V(\partial p/\partial V)_T$, the isometric thermal variance of the pressure, $(\partial p/\partial T)_V$, the isobaric thermal variance of the volume, $(\partial V/\partial T)_p$, and the isobaric specific heat, $C_p$, all calculated at the density at which the system entropy changes sign. All these thermodynamic variables may be observed to satisfy the local conditions of stability in the fluid phase at the Alder-Wainwright transition.

At the Alder-Wainwright transition, the maximum errors in the calculated value of the system entropy are less than 1 part in $10^{-26}$. As a further cross-check upon the internal consistency of the computational procedure, the isobaric thermal variance of the volume, $(\partial V/\partial T)_p$, has been calculated both directly by the formula for the hard-sphere system,

$$\left(\frac{\partial V}{\partial T}\right)_p^{SPT} = \frac{V}{T}\frac{(1-2h^3+h^4)}{(1+4h+4h^2-4h^3+h^4)} = $$
$$= \frac{V}{T}\left[1-\frac{2h(2+2h-h^2)}{(1+4h+4h^2-4h^3+h^4)}\right] \quad (14)$$

and by the thermodynamic equation, $(\partial V/\partial T)_p = -(\partial p/\partial T)_V/(\partial p/\partial V)_T$. The values calculated by the two different methods agree to at least seventeen significant figures.

A less widely used determinant of a phase transition is the isobar inflection, $(\partial Cp/\partial p)_T = 0$. The importance of the isobar inflection for investigating the phase stability of fluids has been clearly described by Bernal[8], who pointed out the experimental demonstration of its significance by Jones and Walker[44], and by Deiters[45]. When the fourth Clausius equation is applied, $(\partial Cp/\partial p)_T = -T(\partial^2 V/\partial T^2)_p$, the isobar inflection can be determined from the second isobaric temperature derivative of the volume. SPT gives for the second isobaric temperature derivative of the volume:

$$\left(\frac{\partial^2 V}{\partial T^2}\right)_p^{SPT} = -\frac{V}{T^2}\frac{(1-h)h^2(1+h+h^2-h^3)(3+9h-6h^2-4h^3+5h^4-h^5)}{(1+4h+4h^2-4h^3+h^4)^3}. \quad (15)$$

The isobar inflection is determined by the roots of the polynomial of eighth degree in the numerator of equation (15). Aside from its obvious roots at zero and unity, equation (15) admits no real roots for values of the free-volume parameter, $h$, greater than zero or less than one; of its four real roots, two are negative and the other two greater than one. (In both equations (14) and (15), the "ideal gas" component of each variable has been factored out in order to provide an immediate sense of the effects of the hard-core component of the intermolecular potential.)

## 6. The temperature and volume limitations of scaled particle theory.

The pressures and reduced volumes of the gas-solid phase transition for a hard-sphere fluid have been calculated at temperatures from 0.01K, to 10,000K. The hard-sphere fluid considered is characterized by a hard-core covolume of 20cm$^3$ and a mass of 35amu; these parameters are typical of light gases such as methane or nitrogen. For comparison, the principal results of the investigation have been carried out also for a gas with covolume and mass five times greater.

Scaled particle theory uses a factored partition function, (5), in which the contribution of the kinetic energy of the fluid has been obtained by integrating out the Maxwell-Boltzmann distribution of kinetic energies of an ideal gas. Therefore, there exist restrictions, of both volume and temperature, upon the validity of the SPT formalism. Specifically, the SPT partition function is restricted such that values of the system vol-ume used satisfy the inequality: $V \gg Nl^3$. In short, the SPT formalism is valid only for such volumes for which the average interparticle separation must be much greater than the particle dimensions.

In Fig. 2, the calculated values of the volume of the hard-sphere gas for which the entropy vanishes, $V(T;S=0)$, are shown on a double-logarithmic scale through the temperature range 0.01K-10,000K, together with the values of its "kinetic" volume, $V_{kin}(T) = Nl^3$. The temperature at which $V = V_{kin}$ is approximately 0.16K, and represents an absolute limit for the validity of the SPT formalism. At such low temperatures, not only has the system volume become very small, in order to provide a zero for the entropy equation, (12), but the thermal de Broglie wavelength has become relatively much larger because of the square root of the temperature in its denominator. The value of temperature at which $V = V_{kin}$ is marked in Fig. 2 by a broad dashed stripe, approximately to divide visually the regions in which the SPT formalsm is valid and invalid. Visual inspection in Fig. 2 of the traces of $V(S=0)$ and $V_{kin}$ persuades that a reasonable limit of confidence for SPT could be placed at about the temperatures 5-10K, for at 10K, $V(S=0) \approx 100V_{kin}$.

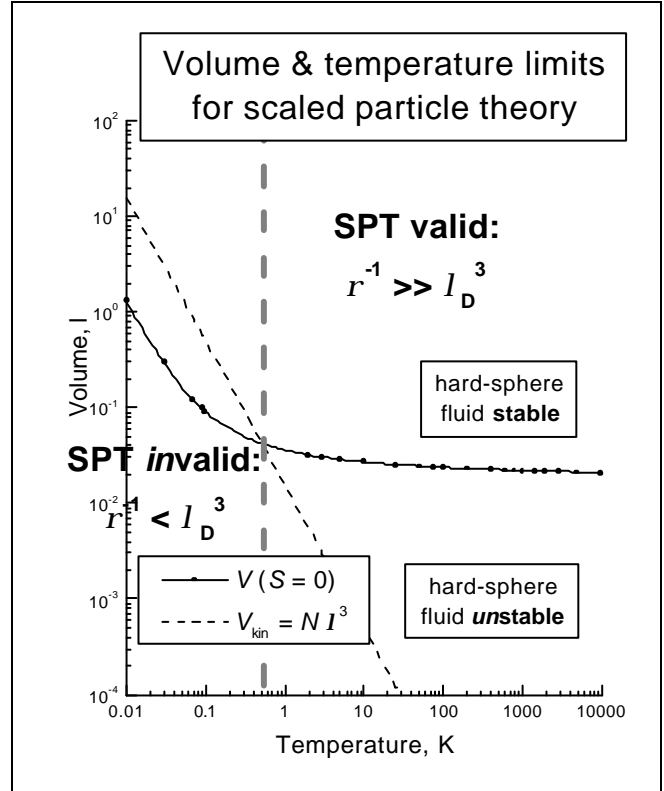

**Fig. 2 Volume and temperature limits of scaled particle theory.**

The pressures at which the entropy of the fluid vanishes, $p(S=0)$, are shown in Fig. 3 on a double-logarithmic scale for the range of temperatures 0.1-10,000K. On Fig. 3, are shown also the values of the pressure for which $V = V_{kin}$, which are marked with a broad dashed stripe, again to indicate the pressure limits of validity of the SPT formalism. The two regions of validity and invalidity, respectively, are indicated in Fig. 3. Shown also on Fig. 3 are the pressures estimated for the phase transition by the fixed value of the reduced volume of the Alder-Wainwright transition given by the value of the free-volume parameter $h^{AW} = 0.67114$. As may be noted in Fig. 3, the predictions of the gas-solid transition pressures from these two calculations give very similar values over a range of temperatures extending three orders of magnitude. At high temperatures, the estimate obtained from the fixed value of $h^{AW}$ underestimates the transition pressure. At low temperatures, the estimates of the transition pressure are plainly too small, and certainly incorrect; for at temperatures less than approximately 500K, a phase transition which might occur at the pressures estimated by $h^{AW}$ would allow the fluid to exist with negative entropy. This error is shown clearly in Fig. 3 where the pressures of the gas-solid phase transition, as determined by the vanishing of the entropy of the fluid, $p(S=0)$, and as estimated by $h^{AW}$, are shown on a double-logarithmic scale which emphasizes the lower temperatures and shows clearly the errors developed by the estimate of a fixed, temperature-independent, value for the free-volume parameter at the gas-liquid phase transition. For temperatures less than 100K, the assumption that the gas-solid phase transition should occur at $h^{AW} = 0.67114$ would allow for the fluid system to exist in a state of negative entropy over ranges of almost an order of magnitude in pressure.

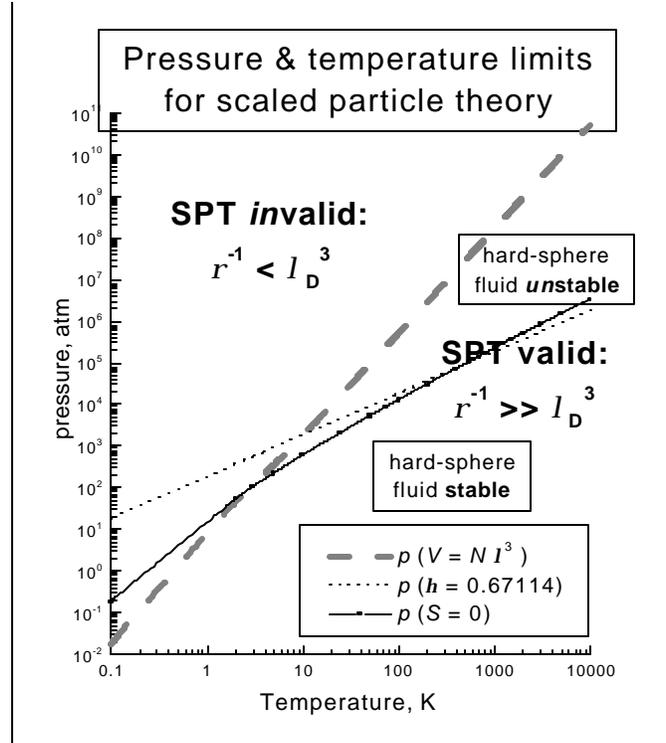

**Fig. 3 Pressure and temperature dependence of the validity of scaled particle theory, and the pressures of the Alder-Wainwright transition, as functions of temperature, $p(S=0;T)$.**

### 7. The high-pressure, supercritical, gas-solid Alder-Wainwright transition and its temperature dependence.

In Fig. 4, are shown the values of the free-volume parameter at the transition point where the entropy vanishes, $h(S = 0)$, as a function of temperature over the range 10K-10,000K. Shown also on Fig. 4 is the trace of the fixed value of the free-volume parameter estimated for the transition point by Alder and Wainwright, $h^{A-W} = 0.67114$. Although the ordinate has been drawn with a very expanded, linear scale in Fig. 4, the value of the free-volume parameter at the transition point determined by the vanishing of the entropy differs by that estimated at 0.67114 by only a few percent: at 100K, $h^{S=0} = 0.63206$, 5.8% lower than $h^{A-W}$; at 1,000K, $h^{S=0} = 0.68227$, 1.6% higher.

# 8. The temperature dependence predicted by SPT for the Alder-Wainwright transition at low temperatures.

A consequence of the third law requires that the specific heat, however measured, must vanish at $T \to 0$ at least as fast as $T^{1+d}$ for $d > 0$. However, as has been noted in section 4, the specific heat for a hard-core fluid described by scaled particle theory remains, at all temperatures, simply that of an ideal gas:

$$C_V^{SPT} = \frac{3}{2} N k_B$$

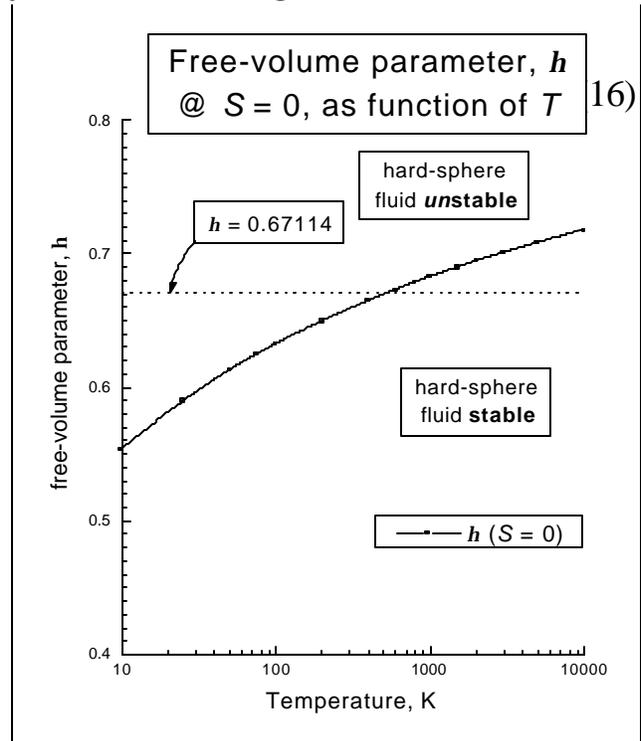

This property, that scaled particle theory describes a fluid phase of which the isometric specific heat fails to manifest correct physical behavior at very low temperatures, suggests that SPT might indicate the presence of a low-temperature gas-solid phase transition. Scaled particle theory does indeed strongly suggest that the fluid undergoes a gas-solid phase transition as its temperature approaches absolute zero.

In Fig. 5, the pressures of the gas-solid phase transition may be observed as a function of temperature over the low range 0.1-15K on a linear scale. While taking cognizance of the temperature limitations of the SPT formalism, specifically that the details of the description of the system are unreliable at temperatures less than approximately 1-5K, the values of the pressure at which the entropy of the fluid becomes zero are nonetheless seen in Fig. 5 to decrease rapidly with temperature and to approach zero as $T \to 0$. Between the temperatures 5-15K, the values of the pressure at which the fluid system becomes absolutely unstable against transition to the solid phase decrease by approximately half an order of magnitude, while the temperature decreases by only 10 degrees. Thus, $p(S=0) \to 0$ as $T \to 0$, and faster than does the temperature itself. The shape of the curve which represents the pressure of the Alder-Wainwright transition at low temperature argues strongly that, as the temperature of the hard-sphere gas approaches absolute zero, the fluid becomes increasingly unstable at any pressure. The pressure determined by $h^{AW}$ is also drawn on Fig. 5, and is easily seen to represent unphysical behavior at low temperatures.

**Fig. 4 Free-volume parameter, $h$, at the transition point determined by $S = 0$, as a function of temperature.**

## 9. The entropy of the gas-solid phase transformation estimated by scaled particle theory.

It has been shown that a hard-sphere system in the fluid state, as described by scaled particle theory, is characterized by limits determined by the stability condition imposed by the third law of thermodynamics. That such limit determines a fluid-solid separation is easily recognized.

The limit set by the change of sign of the system entropy occurs at densities always greater by approximately thirty percent than that of the close-packed configuration. Certainly the hard-sphere system exists at such densities, as well as at greater until that of the close-packed limit. Because scaled particle theory describes the fluid system equally in the dilute gas or dense liquid circumstances, the stability limit must be that for the separation of the system in its fluid solid phases.

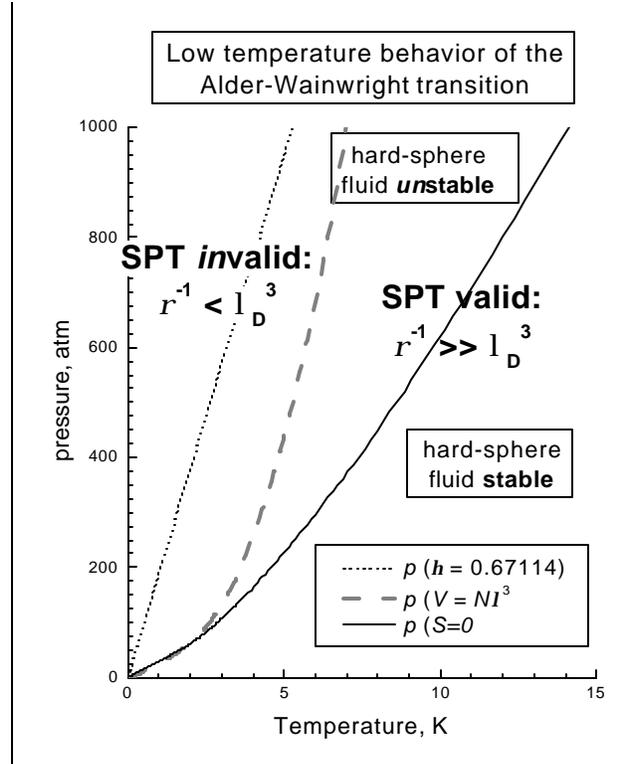

**Fig. 5 Low temperature behavior of the Alder-Wainwright transition predicted by $S = 0$.**

Because SPT predicts a solid-fluid phase transition for a hard-sphere gas, it would be of interest to calculate the latent heat of transformation at its sublimation or melting point. Because a latent heat of transition represents the difference in enthalpy between two phases, the entropy of transformation cannot be calculated immediately for the hard-sphere system because a thermodynamic description is not yet available for the hard-sphere solid. A thermodynamic description of the hard-sphere solid which has been developed within the spirit of scaled particle theory will be offered in a subsequent article; and from such will be given an expression for the entropy of melting at general temperatures. Here is presented an approximate calculation of the limiting value of the SPT entropy of melting at low temperatures.

The entropy of a system which sublimes directly from the solid to the gas phase may be given, at pressure, $p$, and temperature, $T$, by:

$$S(T;p) = \int_0^{T_{sf}} \frac{c_p^s}{T'} dT' + \frac{\Delta L^{sf}}{T_{sf}} + \int_{T_{sf}}^{T} \frac{c_p^f}{T''} dT''. \tag{17}$$

In equation (17), the first term represents the entropy of its solid phase, which exists from absolute zero until the melting temperature, $T_{sl}$; $c_p^s$, $c_p^l$ represent, respectively, the molar isobaric specific heat of the solid and liquid phases; and $\mathbf{D}L^{sl}$ represents the latent heat for the solid-liquid phase transformations. It is emphasized again that equation (17) stands as written, regardless of whatever units used, and that it contains *no* additive constant.

As a hard-sphere system approaches absolute zero, its latent heat of transition for the solid-fluid phase transition, $\mathbf{D}L^{sf}$, approaches a constant:

$$\Delta L(T)^{sf} = \Delta H^{sf} = \left(U(T)^{fluid} - U(T)^{solid}\right) + p^{sf}\left(V(T)^{fluid} - V(T)^{solid}\right), \tag{18}$$

in which $p^{sf}$ represents the pressure of the solid-fluid phase transition. As shown in Fig. 5 of the previous section, the pressure of the fluid-solid phase transition, $p^{sf}$, goes rapidly to zero at low temperatures. Therefore, although there exists a volume difference between the two phases, the second term in parentheses on the right side of equation (18) decreases rapidly as the system temperature approaches absolute zero. The first term in parentheses on the right side of equation (18) does *not* vanish as the system temperature approaches absolute zero. The internal energy of the gas phase vanishes linearly with temperature,

$$U(T)^{fluid} = \frac{3}{2}N k_B T, \tag{19}$$

while that of the solid may be assumed to have the general form

$$U(T)^{solid} \sim \frac{3}{2}N k_B T \left[\frac{2\mathbf{p}^4}{5}\left(\frac{T}{T^D}\right)^3 + \cdots O\left(\exp(-T^D/T)\right)\right], \tag{20}$$

$$\lim\left[\left(\Delta S^{sf}\right)^{SPT} = \frac{\Delta L(T)^{sf}}{T^{sf}}\right] \xrightarrow[T \to 0]{} \frac{3}{2}N k_B \approx 2.98 \text{ cal/moleK}. \tag{21}$$

This interesting property has not been previously reported in the extensive literature dealing with scaled particle theory. The entropy of fusion at the melting point for almost all metals lies in the range 1.7-2.6 cal/moleK.[46] That the entropy of melting for real metals should be less than that given by the SPT limit (21) should be expected, for the limit to which (21) refers is that for which the temperature vanishes. Real metals fuse at high temperatures, at which the calculated entropy would have subtracted from equation (20) contributions for both the binding energy of the metal and also its volume change at melting. The typically high melting temperature of most metals indicates substantial binding energy, and thereby a significant contribution to the change in enthalpy from the term $\Delta U^{sl}/T$. An example of a real material composed of volatile molecules of almost spherical shape, the non-associating hydrocarbon tetramethylmethane, $C_5H_{12}$, melts at 250.2K and possesses a latent heat $\Delta H^{sl} = 778$cal/mole, such that its entropy of transformation $\Delta H^{sl}/T^{sl} = 3.0$cal/degK-mole.[47] Thus, scaled particle theory predicts a limiting value for the transformation entropy which is close to the measured values of real materials.

## 10. Discussion.

The purpose of this short paper has been to demonstrate that scaled particle theory accurately predicts the high-density, gas-solid Alder-Wainwright transition in a hard-sphere system. The qualitative features of the mathematical computations described above resolve all doubts on that question. Furthermore, these results show that SPT additionally describes interesting features of the hard-sphere system not previously been reported: the temperature dependence of the free-volume parameter at the gas-solid phase transition, $h(T)_{S=0}$; and the probable requirement which SPT places upon the hard-core system, as the system temperature approaches absolute zero, to condense into the solid phase at finite temperatures and pressures.

The temperature dependence of the density at which the hard-sphere system undergoes a fluid-solid phase transition, $h(T)_{S=0}$, should be expected. There is an intrinsic ambiguity in the temperature in all computer simulations involving hard-core systems; for such the temperature is always directly proportional to the pressure, and in such as the Clausius diagrams of pressure against volume, the pressure is always given as the "reduced pressure, $p' = p/Nk_BT$. For example, for monte carlo or molecular dynamics calculations, the pressure relationship is given as:[48]

$$\frac{pV}{Nk_BT} - 1 = \frac{pV}{dNk_BT}\frac{1}{t}\sum_{i,j}\int_0^t \vec{r}_{i,j} \cdot \vec{I}_{i,j} d(t'-t) dt = \frac{pV}{dNk_BT}\frac{1}{t}\sum_{i,j} m\Delta\vec{v}_{i,j} \cdot \vec{r}_{i,j}, \qquad (22)$$

where the summations is over all $i$-$j$ collisions within the time limit, $t$, and $r_{i,j}$, is the distance between the centers at the time of collision ($d$ is the dimension of the model). However, the phase rule of Gibbs in no way depends upon any detail of the intermolecular potential; it holds true for a hard-sphere system as any other. A single-component, hard-sphere disk in a single-phase must be described by two independent, intensive variables, and at its binary-phase transition, one.

Although it has been shown that the fluid phase of a hard-core system, as described by scaled particle theory, is characterized by limits determined by the stability condition imposed by the third law of thermodynamics, the detailed quantities calculated above cannot be considered to be strictly accurate, for such were calculated by considering only the entropy of the gas state. An exact calculation of the pressures of the Alder-Wainwright transition must include, in addition to the entropy of the gas phase, both that of the solid phase of the hard-core system and also the entropy of transformation, as set out in equation (17). However, the thermodynamics of the hard-sphere solid have not yet been worked out, and it is therefore not possible to calculate exactly either the entropy of the hard-sphere solid or its melting entropy at general temperatures. As has been pointed out in the previous section, the contribution of the low-temperature limit of the melting entropy is approximately 3cal/moleK, and this value is of the same order as, and very closely approximates, the entire entropy of melting in simple, real materials. Therefore, in order to examine how inclusion of the entropy of melting might alter the details of the Alder-Wainwright transition, as determined by SPT, the pressures for the Alder-Wainwright transition have been calculated using the condition $S(p,T)+(\Delta S)^{sf} = 0$. In Fig. 6, are shown the pressures of the Alder-Wainwright transition as determined by the vanishing of the system entropy as calculated both from the fluid phase alone, and also with inclusion of the contribution of the entropy of melting, as estimated from its low-temperature limit. The data represented in Fig. 6, show clearly that the inclusion of the entropy of melting has no effect upon the qualitative features of the Alder-Wainwright transition and produces only very small quantitative change above the temperature of approximately 100K. Furthermore, the low-temperature behavior of the Alder-Wainwright transition as predicted by SPT, characterized by the vanishing of the fluid phase at all pressures as the temperature goes to zero, remains. The magnitude of the calculated free-volume parameter at the density of the Alder-Wainwright transition as predicted by SPT, $h(S+(\Delta S)^{sf} = 0)$, deserves particular attention: In Fig. 6, it may be observed that the free-volume parameter at the Alder-Wainwright transition is essentially indistinguishable from the fixed value, $h^{AW}=0.67114$, estimated by Alder and Wainwright, over a range of temperatures extending almost three orders of magnitude.

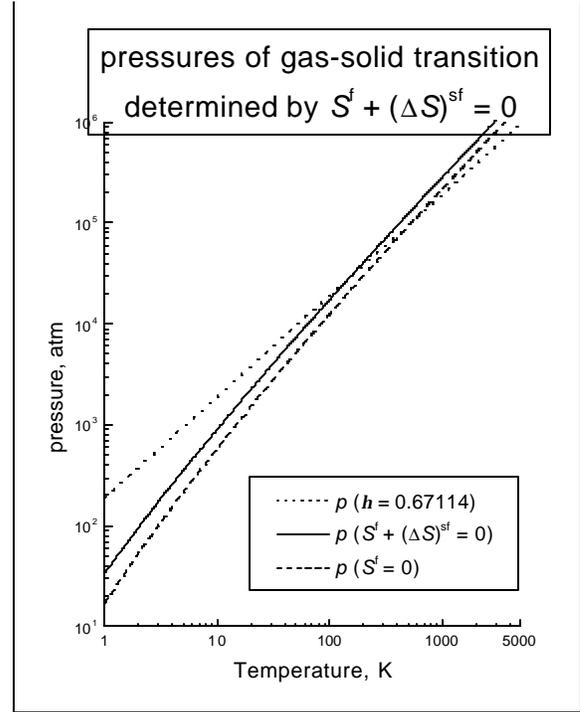

**Fig. 6 Pressure of the Alder-Wainwright transition, determined with inclusion of the entropy of transformation, $(\Delta S)^{gf} \approx 3/2 k_B$.**

Because a density at which the system entropy vanishes determines a point of absolute instability for any phase, the densities of the Alder-Wainwright transition calculated by SPT should be considered the solid-end points of the gas-solid coexistence line for the phase transition, at each temperature. The positions of the gas-end of the gas-solid coexistence line can be determined only with an exact solution of the statistical mechanics of the hard-sphere solid, which does not exist.

This exegesis of the competence of scaled particle theory to predict the high-density, gas-solid phase transition in a hard-core system was occasioned by its previous use in I to describe the high-density, gas-liquid phase transition. Therefore, it seems appropriate to comment upon certain similarities between both analyses. A surprising aspect of the explicit prediction by scaled particle theory of the Alder-Wainwright, fluid-solid phase transition in a hard-sphere system is that it remained so long unrecognized. A similar remark holds for the recognition of the supercritical, high-density, gas-liquid phase transition. A review of the SPT literature on the subject reveals similar causes for both oversights. As pointed out in I, previous investigations seeking the gas-liquid phase transition had involved an unphysical restriction of the state vectors employed in the quantal partition function. Without exception, previous investigators had confined their analyses of hard-sphere systems to partition functions for which the admissible states were, effectively, restricted to sums over gas states. Such restriction "locked in" the results of those analyses to (unsurprisingly) gas states and eliminated the liquid states from consideration, with the consequence that the supercritical, high-density, gas-*liquid* phase transition was inevitably missed. When the partition function for the hard-core system includes real clusters, the supercritical, high-density, gas-liquid phase transition appears naturally. (For an example of the inability to predict a phase transition in a hard-core system which is inevitably consequent of failure to include an adequate set of states in the quantal partition function, or to respect all the requirements for stability, attention is directed to the 1986 article by Speedy.[49]) Similarly, previous investigators of the gas-solid phase transition restricted their considerations of stability conditions of the gas to the local behavior of the isothermal compressibility. When all the requirements of stability are considered, the gas-solid phase transition appears naturally as a consequence of the vanishing of the absolute entropy of the fluid phase.

The description of the high-density Alder-Wainwright transition given above by scaled particle theory is *not* restricted to artificial hard-sphere gases; it holds equally true in all its qualitative features for real fluids. The mathematical description of the entropy of every real fluid system must possess a component which accounts for the entropy of exclusion, as determined by SPT. The entropy of exclusion obtains from the impenetrable property of real molecules, and is a consequence of the mathematics of statistical geometry, and is always *negative*. Furthermore, the *negative* term which describes the entropy of exclusion of a fluid will possess always the second-order singularity determined by scaled particle theory, $S^{hc}(\mathbf{h}) \sim -h(\mathbf{h})/(1-\mathbf{h})^2$, where $h(\mathbf{h})$ represents any non-singular function of $\mathbf{h}$, usually a polynomial. Even when the complications of a non-spherical shape of the hard-core have been taken into account, as has been done by, for examples, Boublik[50, 51], Chapman[52], Nezbeda[53], Vimalchand[54], and Zhou[55], the second-order singularity in the reduced density remains. The same fact holds also when the temperature dependence of the hard core is introduced, as has been done by, for example, Deiters[56, 57]. The entropy of every real gas system is positive at low and modest pressures at all temperatures, of course, for if not, the gas state would always be unstable against transition into the solid state and could not exist.

page 18

However, as stated above, the total entropy of every real gas system possesses also, at all temperatures, a *negative* component, the entropy of exclusion attributable to the universal property of impenetrability of its molecules as described by SPT. The (negative) component, the entropy of exclusion, by virtue of its second-order singularity, is unbounded at high-densities. Therefore, at some sufficiently high density, the fluid phase of every real gas must become absolutely unstable. That instability is the famous Alder-Wainwright transition. The Alder-Wainwright transition is *not* an artifact of the machine-calculated simulations with which Alder and Wainwright first determined the high-density gas-solid phase transition; it is a real phenomenon which occurs in all real fluids.

? Because there are very few models in quantum statistical mechanics which admit exact solutions, the constraints of the third law have not often been applied for stability analysis. Therefore, and in order to show that such analysis does not predict a spurious instability or phase transition, in two Appendices which follow, the identical analysis as has been used for the three-dimensional hard-sphere gas is applied to, respectively, the one-dimensional hard-body, - or "hard rod," – system, and the two-dimensional, - or "hard-disk," - system. In those Appendices is shown that, the constraints of the third law, when applied to scaled particle theory and the one- and two-dimensional hard-core systems, do not predict a spurious instability for the first case, and do predict one-such, and its accompanying first-order, fluid-solid phase transition, in the second case, exactly as demonstrated by molecular dynamics computer simulations.[58-60]

Notwithstanding its use of the modern formalism of scaled particle theory, this prediction of the Alder-Wainwright, fluid-solid phase transition, above, is very much in the spirit of the analysis of the gas-liquid phase transition by J. D. van der Waals. The traditional prediction of the low-temperature gas-liquid phase transition obtained by the van der Waals equation of state does not strictly describe any liquid configuration. When the van der Waals equation of state is used to describe a gas, its isothermal compressibility is observed to change sign at a certain (critical) temperature; and for all lower temperatures, the gas-liquid phase coexistence line is inferred *a fortiori*, using the (appropriately named) "Maxwell construction." Similarly, when scaled particle theory is used to describe a hard-sphere gas, its entropy is observed to change sign at a certain density; and for all higher densities, the solid state is inferred *a fortiori*.

**Appendix A.: Application of the constraints of the third law to the one-dimensional hard-rod system:**

The statistical mechanical problem of a one-dimensional hard-body gas admits an exact solution, and *no* phase transition. The equation of state of a one-dimensional hard-body fluid is given by the Tonks equation:[61]

$$p^{\text{SPT-1dim}} = \frac{Nk_BT}{L}\frac{1}{(1-h)} = Nk_BT\left[\frac{1}{L} + \frac{h}{L(1-h)}\right] = p^{\text{IG}} + p^{\text{hc}} \tag{23}$$

where $h = Ns/L$, in which $s$ represents the length of the hard rod and $L$ the system dimension ("volume"). The expression for the pressure has been factored in equation (23) in order to separate the respective "ideal gas" and hard-core contributions to the pressure. The Helmholtz free energy of the one-dimensional hard-rod system is given explicitly as:

$$F^{\text{SPT-1dim}} = -Nk_BT\left[\ln\left(\frac{L}{Nl}\right) + 1 - \ln(1-h)\right]. \tag{24}$$

The hard-core terms in the free energy, and the canonical partition function, of a hard-rod system in one dimension are purely entropic, as for the three-dimensional case; and the internal energy, $U$, of the hard-rod system is exactly $1/2 Nk_BT$. The entropy of the hard-rod system is therefore given explicitly by $S = (U-F)/T$ as:

$$S^{\text{SPT-1dim}} = Nk_B\left[\frac{3}{2} + \ln\left(\frac{L}{Nl}\right) + \ln(1-h)\right], \tag{25}$$

in which the thermal de Broglie wavelength, $l = h/\sqrt{(2\pi k_B T)}$.

The condition of validity for the statistical mechanical solution of the one-dimensional hard-rod system given by equation (23) is analogous to the same for the hard-sphere system in three dimensions:

$$\frac{Nl}{L} \ll 1. \tag{26}$$

If the one-dimensional hard-rod system were compressed to the length at which the entropy, as given by equation (25), would vanish, its length would then have to satisfy the equation:

$$\frac{Nl}{L_{S=0}} = \exp\left(\frac{3}{2}\right)(1-h) \approx 4.482(1-h). \tag{27}$$

However, both equations (26) and (27) cannot be satisfied simultaneously. Because the limits of the free-volume parameter are $0 \leq h \leq 1$, the value of the system length at which the expression for the entropy, as given by equation (25), vanishes, $L_{S=0}$, clearly lies outside the limits of validity, given by (26), of the equation of state and the hard-rod partition function, from which equation (25) obtained. (The mathematically valid but physically meaningless values of, say, $h = (0.9999999\cdots)$, are not considered, for such must be interpreted simply as the system compressed to close-packing, not as a phase transition. For all fluid-solid phase transitions, the change in density at the temperature of transformation is approximately 10-15%. Therefore, when considering equation (27), it would be reasonable to restrict the free-volume parameter such that $h \leq \sim 0.9$.)

Therefore, the stability condition that the entropy of the system must always be positive does *not* predict any phase transition for the *one*-dimensional hard-core system at any volume within the limits of validity of its description. Application of the constraints of the third law to the one-dimensional system does *not* involve any prediction of instability or phase transition for the one-dimensional hard-rod gas, and is consistent with other stability analyses.

**Appendix B.:  Application of the constraints of the third law to the two-dimensional, hard-disk system:**

? Scaled particle theory develops the exact solution also for the equation of state of the two-dimensional hard-body system, - or "hard-disk" system:

$$p^{\text{SPT-2dim}} = \frac{Nk_BT}{A}\frac{1}{(1-h)^2} = Nk_BT\left[\frac{1}{A} + \frac{h(2-h)}{A(1-h)^2}\right] = p^{\text{IG}} + p^{\text{hc}}$$

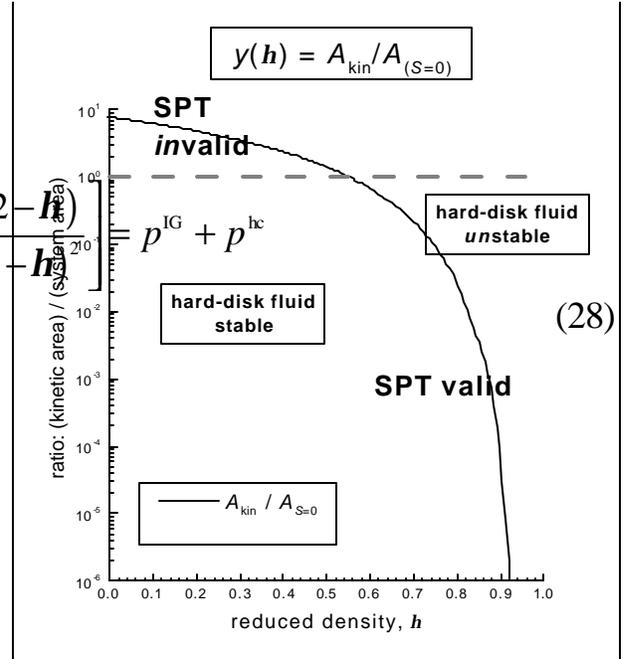

(28)

where $h = r\pi s^2/4$, in which $s$ represents the diameter of the hard disk and $A$ the system area ("volume"). The expression for the pressure in (28) has been factored in order to separate the respective "ideal gas," $p^{\text{IG}}$ and hard-core contributions, $p^{\text{hc}}$. The Helmholtz free energy of the two-dimensional hard-disk system is given explicitly as:

**Fig. 7 Ratio of kinetic area to system area of a two-dimensional hard-disk system at the point $S = 0$, as a function of temperature.**

$$F^{\text{SPT-2dim}} = Nk_BT\left[-\ln\left(\frac{A}{A_{\text{kin}}}\right) - 1 + \frac{h}{(1-h)} - \ln(1-h)\right]$$

. (29)

In deriving equation (28), the ideal gas factor in the partition function which involves the kinetic volume, $V_{\text{kin}}$, has been written as $A_{\text{kin}}$. This term is formally identical for a one-, two-, or three-dimensional system, and may be written as:

$$V_{\text{kin}}^{\text{D-dim}} = (N l^3)^{D/3}, \tag{30}$$

in which $l = h/\sqrt{(2\pi k_B T)}$, the thermal de Broglie wavelength. Thus, for a two-dimensional, hard-disk fluid, its kinetic volume is $A_{\text{kin}} = Nd^2$, where $(Nd)^{2\text{-dim}} = (N^{3\text{-dim}})^{2/3}$. In integrating out equation (29), the boundary condition has been used that the hard-core component of the Helmholtz free energy $F^{\text{hc}} \to 0$ as $V \to \infty$ (i.e., as $h \to 0$). The hard-core terms in the free energy, and the canonical partition function, of a hard-disk system in two dimensions are purely entropic, as for the three-dimensional case; and the internal energy, $U$, of the hard-disk system is exactly $Nk_BT$. The entropy of the hard-disk system is therefore given explicitly by $S = (U-F)/T$ as:

$$S^{\text{SPT-2dim}} = Nk_B \left[ 2 + \ln\left(\frac{A}{A_{\text{kin}}}\right) - \frac{h}{(1-h)} + \ln(1-h) \right]. \tag{31}$$

The condition of validity for the statistical mechanical solution of the two-dimensional hard-disk system described by equation (28) is:

$$\frac{A_{\text{kin}}}{A} \ll 1. \tag{32}$$

(Because of formal equivalence of derivation for a hard-core gas in one, two, or three dimensions, the condition of validity of the hard-core fluid in any dimension, $D$, could be written as $(Nl^3)^{D/3}/L^D \ll 1$.) If the two-dimensional hard-disk system were compressed to the area at which the entropy, as given by equation (31), would vanish, its area would then have to satisfy the equation:

$$\frac{A_{\text{kin}}}{A_{S=0}} = y(h) = (1-h)\exp\left(2 - \frac{h}{(1-h)}\right) \approx 7.39(1-h)\exp\left(-\frac{h}{(1-h)}\right). \tag{33}$$

At low densities, equations (32) and (33) obviously cannot be satisfied simultaneously. However, the limits of the free-volume parameter, or reduced density, are $0 \leq h \leq 1$, and the value of the function $y(h)$ diminishes rapidly for values of its argument greater than approximately 6. A plot of $y(h)$ is shown in Fig. 7 for values of the argument between $10^{-4}$-0.95, in which can be seen that the requirements of inequality (32) are easily satisfied for reduced densities greater than 6. The line $A_{\text{kin}}/A = 1$, which marks the absolute limits of validity of the SPT formalism is also shown as a broad, dashed gray band on Fig. 7.

The region within which the gas-solid phase transition occurs is to the right of the trace of $y(h)$ and (well) below the line $A_{\text{kin}}/A = 1$. The limits of the validity of the SPT formalism in two dimensions are shown graphically also in Fig. 8 for a two-dimensional gas of particles characterized by a mass of 35 amu and a molar co-volume of 20 cm$^3$, which values are typical for a light gas. In Fig. 9 is shown graphically the value of the free-volume parameter at its values for which the system entropy of the fluid phase vanishes, $h(S = 0)$. Similarly as a real, three-dimensional system, the transition value of the free-volume parameter depends weakly upon temperature. At 100 K, $h^T = 0.793$;
at 1,000 K, $h^T = 0.890$ K. For reference, the value of the free-volume parameter at close-packing, $h^{c-p} = 2/\sqrt{3} \approx 1.15$ has also been plotted in Fig. 9.

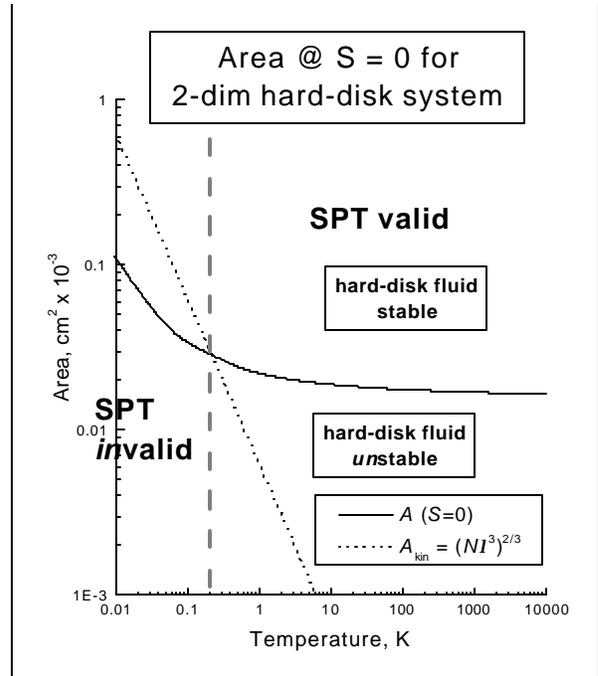

**Fig. 8 Area of the two-dimensional hard-disk system at the point $S = 0$, and its kinetic area, $A_{\text{kin}} = (Nl^3)^{2/3}$, as functions of temperature.**

Therefore, as may be noted in Fig. 9, when the constraints of the third law of thermodynamics are applied, scaled particle theory unambiguously establishes an absolute instability of the fluid phase in a two-dimensional system at temperatures greater than 10 K for values of the free-volume parameter greater than approximately 0.80-0.88, depending upon the temperature. The fluid-solid phase transition in two-dimensional hard-disk systems, at exactly these densities, has already been demonstrated (most recently) by Mitus, Weber, and Mark,[58] and (two decades ago) by Hoover, Hoover, and Hanson.[59]

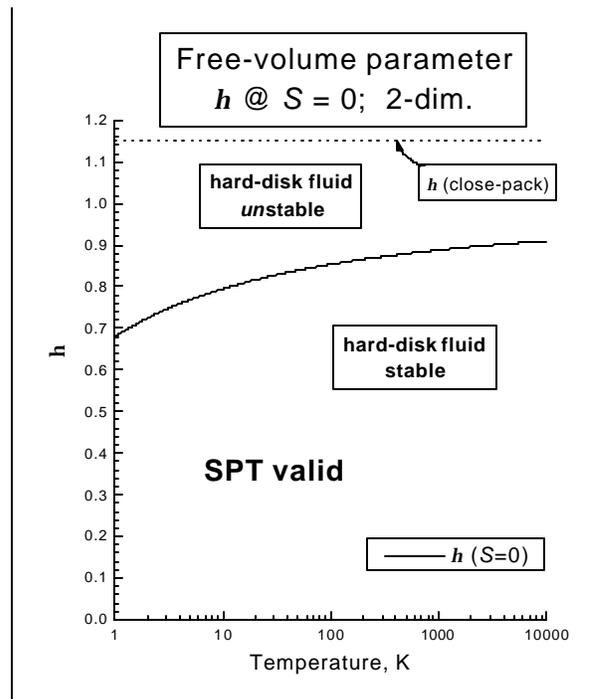

**Fig. 9 Free-volume parameter at the transition point, $h(S = 0)$, of the two-dimensional hard-disk system as a function of temperature.**

**Acknowledgments.**
The author expresses specific gratitude to his colleagues of the Russian Academy of Sciences at the Joint Institute of the Physics of the Earth for their encouragement, their critical review, and their unstinting generosity in offering guidance and detailed knowledge of the many facets of the physics of multicomponent systems in regimes of high density. Special thanks go also to Professor Ulrich K. Deiters of the University of Cologne for reviewing, criticizing, and encouraging publication of this paper.

The author acknowledges with gratitude also the useful discussions with Professor John W. D. Connolly, Director of the National Supercomputing Center at the University of Kentucky concerning the essential ambiguity of temperature in hard-core molecular dynamics and Monte Carlo computer simulations.

In the section for Acknowledgements in the paper I, there is an omission. Before submission for publication, that paper had been thoroughly reviewed by Professor

Ilya Prigogine of the Free University Brussels and the University of Texas. Professor Prigogine's extensive review, incisive analysis, and many questions assisted to focus and refine the author's presentation. When paper I was submitted for publication, Professor Prigogine's name was omitted, as a matter of professional discretion. That omission is herewith corrected; the author thanks Professor Prigogine for his assistance and courtesy.

This work was supported by the Russian Academy of Sciences and, in part, by research grant 8-187-2 from Gas Resources Corporation, Houston.
**List of Symbols:**

| | |
|---|---|
| $A$ | thermodynamic Affinity |
| $c$ | Prigogine $c$-factor |
| h | Planck's constant |
| $k_B$ | Boltzmann's constant |
| $m$ | molecular mass |
| $N_A$ | Avagadro's constant |
| $N$ | number of molecules |
| $n$ | number of moles |
| $p$ | pressure |
| $Q$ | canonical partition function |
| R | universal gas constant |
| $T$ | absolute temperature |
| $V$ | volume |

Greek letters

| | |
|---|---|
| $\beta$ | Boltzmann factor, $= (k_B T)^{-1}$ |
| $\lambda$ | thermal de Broglie wavelength |
| $\eta$ | SPT free-volume parameter |
| $\mu$ | chemical potential |
| $\tau$ | close-packing parameter, $= \sqrt{2}\pi/6$ |

Superscripts

| | |
|---|---|
| SPT | Scaled Particle Theory |
| IG | ideal gas |
| hc | hard core |
| $j$ | phase |
| sl | solid-liquid |
| lg | liquid-gas |

page 24

Subscripts
c   critical
*i*  component